\documentclass[amssymb,prb,twocolumn,floats,amsmath,showpacs,superscriptaddress]{revtex4}
\usepackage[T1]{fontenc}
\usepackage{bm}
\usepackage{graphicx}% Include figure files
\usepackage{amssymb}
\usepackage{amsfonts}
\usepackage{amsmath}
\usepackage{epstopdf}
\usepackage{color}

\newcommand{\deltaE}{\Delta \mathrm{E_{X - XX}}}

\def \FUW{Institute of Experimental Physics, Faculty of Physics, University of Warsaw, Pasteura 5 St., 02-093 Warsaw, Poland}

\def \INP{Institute of Nuclear Physics, Polish Academy of Sciences, Radzikowskiego 152 St., 31-342 Krak\'{o}w, Poland}

\begin{document}

\title{Effect of electron-hole separation on optical properties of individual Cd(Se,Te) Quantum Dots}
%\title{Fluctuation type Cd(Se,Te) Quantum Dots in ZnSe barrier: Beyond the Crossroad of Se and Te Based Systems}

\author{M. \surname{\'{S}ciesiek}}\affiliation{\FUW}
\author{J. \surname{Suf\mbox{}fczy\'{n}ski}}\email{Jan.Suffczynski@fuw.edu.pl}\affiliation{\FUW}
\author{W.~\surname{Pacuski}}\affiliation{\FUW}
\author{M.~\surname{Parli\'{n}ska--Wojtan}}\affiliation{\INP}
\author{T. \surname{Smole\'{n}ski}}\affiliation{\FUW}
\author{P.~\surname{Kossacki}}\affiliation{\FUW}
\author{A.~\surname{Golnik}}\affiliation{\FUW}

\date{\today}

%\keywords{II-VI semiconductors, Photoluminescence, Quantum Dots, exciton in magnetic field}

%################################################################# ABSTRACT

\begin{abstract}

Cd(Se,Te) Quantum Dots (QD) in ZnSe barrier typically exhibit a very high spectral density, which precludes investigation of single dot photoluminescence. We design, grow and study individual Cd(Se,Te)/ZnSe QDs of low spectral density of emission lines achieved by implementation of a Mn-assisted epitaxial growth. We find an unusually large variation of exciton-biexciton energy difference (3 meV $\leq$ $\deltaE$ $\leq$ 26~meV) and of exciton radiative recombination rate in the statistics of QDs. We observe a strong correlation between the exciton-biexciton energy difference, exciton recombination rate, splitting between dark and bright exciton, and additionally the exciton fine structure splitting $\delta_1$ and Land{\'e} factor. Above results indicate that values of the $\delta_1$ and of the Land{\'e} factor in the studied QDs are dictated primarily by the electron and hole respective spatial shift and wavefunctions overlap, which vary from dot to dot due to a different degree of localization of electrons and holes in, respectively, CdSe and CdTe rich QD regions.
\end{abstract}
%####################################

\maketitle

%\thispagestyle{empty}

%#################################### INTRODUCTION
\section{Introduction}
Three dimensional quantum confinement of electrons (\emph{e}) and holes (\emph{h}), such as in a semiconductor Quantum Dot (QD) potential, results in a variety of effects resulting from a discrete energy levels structure and an increased overlap between the \emph{e} and \emph{h}. Typically, a QD and a surrounding barrier are made of materials sharing a common cation, to recall example of InAs/GaAs, CdTe/ZnTe or CdSe/ZnSe QDs. The QDs for which the QD and the barrier materials have no common atom are by far less investigated, despite of several advantages they offer. In the case of such QDs with type-I confinement one should mention here larger band offsets providing higher confining potentials for both, \emph{e} and \emph{h}. Such mixed structures may lead also to type-II confinement, with a decreased wavefunction overlap of the confined \emph{e} and \emph{h}. Such increased \emph{e-h} spatial separation and resulting enhancement of confined carriers lifetimes is highly attractive for studies of magnetic polaron in semimagnetic QDs\cite{Klopotowski:PRB2011, Barman:PRB2015} or Aharonov-Bohm effect,\cite{Sellers:PRL2008} as well as for implementation of QDs in photovoltaic devices\cite{Boyer:NRL2012, Bragar:JAP2012, Dhomkar:SEMSC2013}. The intermediate case between type-I and type-II confinement, as in the present study, offers a chance of investigation of the impact of \emph{e} and \emph{h} wavefunctions overlap on properties of excitonic QDs emission.

\begin{figure}[t!]
\centering
\includegraphics[width = 87 mm]{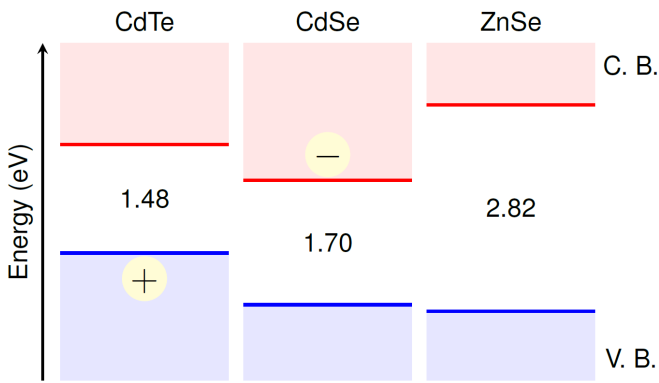}
\caption{Band gap energies and relative band offsets for CdTe, CdSe and ZnSe.\cite{Hsu:JVSTB1988, Hinuma:PRB2014}}
\label{fig:bands}
\end{figure}

In the present work, we design, produce and study individual Cd(Se,Te) QDs embedded in ZnSe barrier. Any composition gradient within the Cd(Se,Te) QD volume results in a decreased spatial overlap of wavefunctions of the \emph{e} and \emph{h}, due to localization of the \emph{e} and \emph{h} in regions, which are richer, respectively, in CdTe and CdSe (see Figure~\ref{fig:bands} providing a scheme of respective band alignments).

We implement a Mn-assisted epitaxial growth, which allows us to obtain a very low spectral density of the QDs, which enables us to study emission of single QDs. We find a wide distribution of exciton (X) and biexciton (XX) emission energy difference (3 meV $\leq$ $\deltaE$  $\leq$ 26~meV). A large scatter is found also for the X lifetime (280 - 620~ps). A statistics collected on several QDs indicates that the $\deltaE$ is strongly correlated with the X lifetime and a bright-dark exciton splitting $\delta_0$, both quantities being a direct measure of \emph{e-h} wavefunctions overlap.\cite{Franceschetti:PRB1998, Takagahara:PRB2000, Bayer:PRB2002} In addition, we establish that the $\deltaE$ is strongly correlated with such QD emission properties as the magnitude of the X fine structure exchange splitting $\delta_1$ and Land{\'e}-factor. This indicates a way for a straightforward identification of QDs with properties desired for a given implementation\cite{Jakubczyk:ACSNano2014, Pacuski:CGrowthDes2014} without a need of in-depth studies, like emission dynamics or magnetophotoluminescence. The reported distinct dependence of the X fine structure splittings and Land{\'e} factor on a degree of \emph{e-h} wavefunction overlap has been rarely observed in II-VI QD systems so far, despite that it was well established in the case of III-V low dimensional heterostructures\cite{Seguin:PRL2005, Pryor:PRL2006, Kleemens:PRB2009, Krapek:PRB2015}. Here, we benefit from the fact that, in contrary to the case of typical binary II-VI QD systems, the \emph{e-h} separation changes significantly from QD to QD, enabling us to perform a systematic study.

%#################################################################
\section{Experiment}
For micro-Photoluminescence ($\mu$-PL) studies the sample is placed in a pumped helium cryostat (temperature down to $T = 1.5$~K) equipped with a superconducting split coil magnet. A magnetic field of up to $B = 10$~T is applied either in Faraday ($\vec{k} \parallel \vec{B}$) or Voigt ($\vec{k} \perp \vec{B}$) configuration. The magnetic field is perpendicular or parallel to the sample surface in this two respective cases. The QDs emission is excited above the barrier bandgap either in a continuous-wave mode at E$_{exc}$ = 3.06~eV ($\lambda_{exc}$ = 405~nm) or in a pulsed mode (100 fs pulse duration, 75~MHz repetition rate) at 3.26~eV ($\lambda_{exc}$ = 380~nm). The excitation beam is focused on the sample surface down to 0.5~$\mu$m diameter spot with an immersion type microscope objective\cite{Jasny:RevSciInstr1996} (NA = 0.72). In the time-integrated measurement the signal is detected using a CCD camera coupled to a 0.75~m spectrometer (50~$\mu$eV of overall spectral resolution). Linear and circular polarizations of the signal are resolved. In the time-resolved $\mu$-PL measurements, the signal is detected by a Hamamatsu Photonics C5680 streak camera with 5~ps temporal resolution.

The scanning transmission electron microscope (STEM) micrographs of the samples are obtained in high-angle annular dark field detector using FEI Tecnai Osiris instrument. The analysis is carried out at an acceleration voltage of 200~keV and point resolution of $\sim$1.36 \AA. In order to determine the concentration of Zn, Se, Te and Cd in the structure cross-section, a local chemical analysis by energy-dispersive X-ray spectroscopy (EDX) is performed on the 100~nm thick specimen. The TEM specimen is prepared using the classical method of mechanical polishing followed by ion milling down to electron transparency.

%#################################################################
\section{\label{sec:samples} Cd(Se,Te)/ZnSe Quantum Dots produced by a Mn-assisted epitaxial growth}

So far, mixed, Se and Te based, QDs were achieved primarily by a colloidal synthesis\cite{Kim:JACS2003, Bailey:JACS2003, Smith:NatureNano2008, Dorfs:Small2008, Hewa-Kasakarage:JPC2009, He:PRL2010, Li:SciRep2014}. However, integration of such colloidal QDs into operational devices might be challenging. It should be easier in the case of epitaxially grown structures, which offer also a feasibility of p- and n- type doping. On the other hand, strain resulting from a very large lattice mismatch\cite{Konagai:JCGrowth1988, Gil:JCGrowth1994, Rubini:JVSTB2000, Toropov:APL2006, Sedova:Semicon2007} between CdTe (a$_{\mathrm{CdTe}}$ = 0.648~nm) and ZnSe (a$_{\mathrm{ZnSe}}$ = 0.567~nm) (above 14\%) leads to a high spatial density of QDs combining these two compounds. This has precluded spectroscopy of individual QDs in such mixed type systems so far.\cite{Toropov:APL2006, Kuo:APL2006, Yang:Nanotechnol2007, Sedova:Semicon2007, Gong:PRB2008, Sellers:PRL2008, Barman:PRB2015, Fan:JCGrowth2015}

\begin{figure}[b!]
\centering
\includegraphics[width= 95 mm]{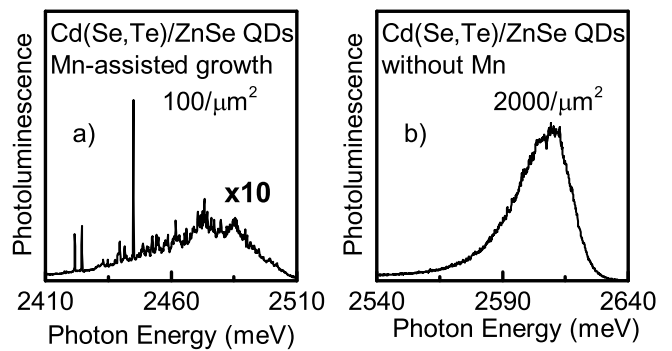}
\caption{$\mu$-Photoluminescence spectra of: a) Cd(Se,Te)/ZnSe QDs produced by Mn-assisted growth (note that the signal intensity is multiplied by 10), b) reference Cd(Se,Te)/ZnSe QDs. Much decreased spectral density of QD emission lines is evidenced in a). The estimated QDs spatial densities are provided in each panel.}\label{fig:gestosc_vs_Mn}
\end{figure}

Cd(Se,Te)/ZnSe QDs presented in this letter are grown by molecular beam epitaxy on a 1~$\mu$m thick ZnSe buffer deposited on a GaAs substrate. They form in a self-assembled mode out of atomic layers of Se, Cd, Te, Cd, and Se (each one for time of 15~sec) consecutively deposited at temperature of the substrate set to 335~${}^\circ$C. An approximately monolayer thick CdSe layer (a$_{CdSe}$ = 0.608~nm) inserted between CdTe layer and ZnSe diminishes the strain resulting from the CdTe and the ZnSe lattice mismatch.\cite{Toropov:APL2006, Sedova:Semicon2007} %The average molar density of the Te atoms in the Cd(Se,Te) QDs is determined to $x = 0.22 \pm 0.03$ using a method presented in the Appendix.
A small flux of the Mn ions assists deposition of the Te layer. The QD layer is covered with a 80~nm thick ZnSe cap. A reference sample, for which no Mn ions are introduced is also grown under the same conditions.

The presence of Mn induces a significant decrease of spectral density of QD emission lines (see Fig.~\ref{fig:gestosc_vs_Mn}). A spatial density of emitting QDs estimated from $\mu$-PL spectra drops from 2000/$\mu$m$^2$ in the case of the reference sample down to 100/$\mu$m$^2$ in the case of the sample produced by the Mn-assited growth. The reduction of QD density for over an order of magnitude results most likely from a reduced mobility of adatoms on the sample surface during the QD nucleation when the Mn atoms are present.\cite{Kim:JCG2000} The energy of QDs emission lowered with respect to the reference sample case (Fig.~\ref{fig:gestosc_vs_Mn}) suggests that the presence of Mn induces also an increased QD size. A qualitatively similar, strong effect of Mn incorporation on the surface morphology and plastic relaxation has been recently reported on also for (Al,Ga)N/GaN heterostructures.\cite{Devillers:CrystGDes2015}

\begin{figure*}[!t]
\centering
\includegraphics[width= 177 mm]{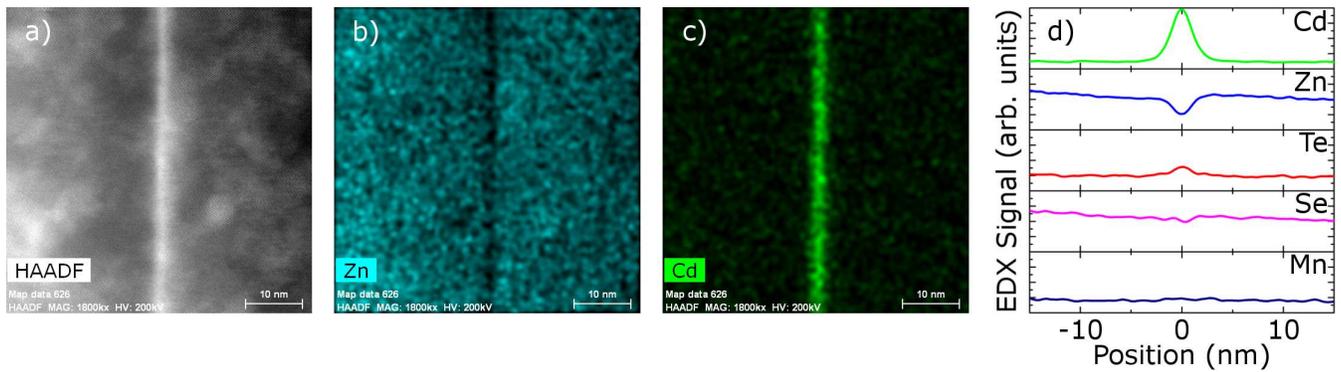}
\caption{a) Scanning transmission electron microscope image with the clearly visible layer of Cd(Se,Te)/ZnSe QDs. EDX maps revealing an increase of Cd content (b) and a decrease of Zn content (c) in the QD layer. d) Vertically integrated EDX signal \textsl{vs} the distance from the QDs plane for Cd, Zn, Te, Se and Mn.} \label{fig:HAADF}
\end{figure*}

The HAADF micrographs confirm the presence of a QD layer in the studied structures (see Fig.~\ref{fig:HAADF}a)). Chemical analysis involving the EDX spectroscopy reveals an expected decrease of the signal related to Zn (Fig.~\ref{fig:HAADF}b) and an increase of the signal related to Cd (Fig.~\ref{fig:HAADF}c)) in the QD layer. Not complete vanishing of the signal related to the Zn in the QD layer (Fig.~\ref{fig:HAADF}b)) is understood when taking into account that the specimen thickness (100~nm) is larger than an individual QD diameter. In such a case signals from both, the Cd(Se,Te) QDs and the ZnSe barrier, contribute to the EDX spectra in the QD layer spatial region.

Figure~\ref{fig:HAADF}d) shows the vertically integrated EDX signal for Zn, Se, Cd, Te, Mn as a function of the distance from the QDs plane. It confirms in a direct way the intended presence of the Se and Te atoms in the QDs layer. The average tellurium content in the QDs is determined to $x_{\mathrm{Te}} = 0.22 \pm 0.03$ using a procedure described in the Appendix. Note that the Mn concentration is too low to be detected by the chemical analysis.

%#################################################################
\section{\label{sec:Results} Results}

\subsection{\label{sec:TRPL} Impact of \emph{e-h} wavefunctions overlap on confined exciton dynamics and isotropic part of \emph{e-h} exchange interaction}

\begin{figure*}[!t]
\centering
\includegraphics[width= 180 mm]{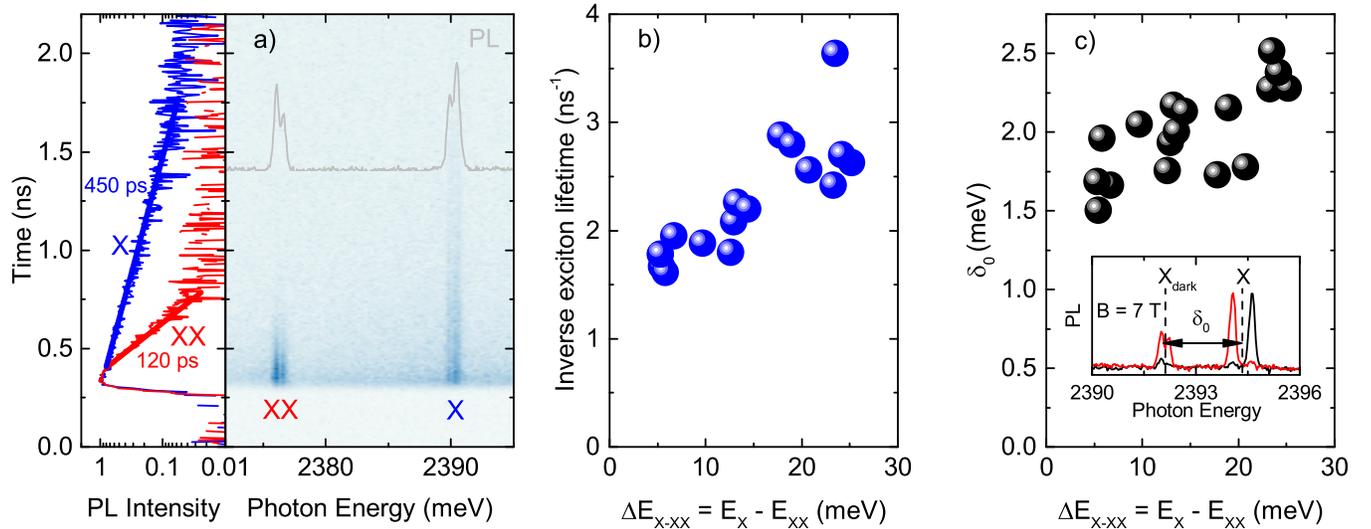}
\caption{a) Streak camera image with the emission intensity profile as a function of time (left panel) and of energy for a single Cd(Se,Te)/ZnSe QD. b) Decay rate of the X as a function of energy difference $\deltaE$ between the X and the XX. c) Splitting $\delta_0$ between bright and dark exciton transitions as a function of the $\deltaE$. Inset: The bright and dark (X$_d$) exciton transitions of an example QD in magnetic field applied parallel to the sample plane resolved in orthogonal linear polarizations.} \label{fig:tau}
\end{figure*}

A temporal evolution of the emission spectrum of a selected Cd(Se,Te)/ZnSe QD following the excitation pulse is presented in an image acquired by the streak camera (Fig.~\ref{fig:tau}a)). Excitonic transitions shown are identified as the recombination of the X and XX confined in the same QD.\cite{Kudelski:JofLumin2005} A doublet of lines evidenced in the case of both, the X and XX, reflects a fine structure exchange splitting $\delta_1$ characteristic for a neutral exciton confined in an anisotropic QD.\cite{Kulakovskii:PRL1999, Bayer:PRB2002}

Emission lifetime at the energy of the X and XX transitions is determined by fitting of a monoexponential decay curve to the emission intensity at respective cross-sections of the streak camera image (see Fig.~\ref{fig:tau}a)). The X lifetime determined for the selected QD equals to 450~ps. In the QD statistics, the X lifetime ranges from 620~ps, which is rather long for II-VI epitaxial QDs, down to 280~ps, in consistency with the previous reports on CdSe/ZnSe QDs.\cite{Patton:PRB2003} The resulting decay rates $\Gamma_\mathrm{X}$ determined as an inverse of the lifetime are plotted against the X-XX energy difference $\deltaE$ in Fig.~\ref{fig:tau}b). It is seen that the $\Gamma_\mathrm{X}$ significantly increases with the increasing $\deltaE$. We note that the biexciton decay time of 120~ps corresponds to the decay rate ($\Gamma_\mathrm{XX}$) around 9~ns$^{-1}$, that is up to 6 times faster than the $\Gamma_\mathrm{X}$, much more than observed previously for the pure CdSe/ZnSe QDs.\cite{Bacher:PRL1999, Patton:PRB2003} The $\Gamma_\mathrm{XX}$ only slightly increases with the $\deltaE$ (not shown).
%This points toward more symmetric spatial distribution of the wavefunctions of the \emph{e} and \emph{h} within the second confined \emph{e-h} pair.

In order to asses factors governing the X lifetime, we determine the bright-dark exciton splitting $\delta_0$ resulting from isotropic component of the \emph{e-h} exchange interaction. The $\delta_0$ has been shown to be a direct measure of the \emph{e-h} wavefunctions overlap.\cite{Takagahara:PRB1993, Takagahara:PRB2000, Bayer:PRB2002} To determine the $\delta_0$ we brighten the dark exciton state X$_d$ by application of the magnetic field in a direction parallel to the sample plane (see the inset to Fig.~\ref{fig:tau}c)).\cite{Nirmal:PRL1995} We trace the energy position of the X$_d$ as a function of the magnetic field and we determine its energy in the limit of zero magnetic field by extrapolation. Values of the $\delta_0$ of the order of 1~meV are found, as expected for II-VI QDs.\cite{Puls:PRB1999, Kulakovskii:PRL1999, Patton:PRB2003} We find that the $\delta_0$ increases with the increasing $\deltaE$ (see Fig.~\ref{fig:tau}c)). Comparison of Figs.~\ref{fig:tau}b) and c) reveals a high degree of correlation between the exciton radiative decay rate and the $\delta_0$. This confirms that: (i) dominating contribution to the $\delta_0$ comes from a short-range component of the \emph{e-h} exchange interaction within the exciton, in agreement with previous theoretical considerations,\cite{Takagahara:PRB1993, Nirmal:PRL1995, Takagahara:PRB2000, Bayer:PRB2002} (ii) the X decay in the studied QDs is governed by a radiative recombination, while non-radiative processes are negligible.

A large range of variation of the $\deltaE$, $\Gamma_\mathrm{X}$ and $\delta_0$ suggests that the \emph{e-h} spatial separation changes strongly from dot to dot. The increase of the exciton radiative decay rate and of the $\delta_0$ with the increasing $\deltaE$ allows us to treat the $\deltaE$ as an estimate of the degree of \emph{e-h} wavefunctions overlap in the further discussion. Time integrated measurements presented in Sec.~\ref{sec:IntPL} shed more light on optical properties of the studied QDs.

%%%%%%%%%%%%%%%%%%%%%%%%%%%%%%%%%%%%%%%%%%%%%%%%%%%%%%%%%%%%%%%%%%%%%%%%%%%%%%%%%%
\subsection{\label{sec:IntPL} Impact of \emph{e-h} wavefunctions overlap on Land{\'e} factor and anisotropic part of \emph{e-h} exchange interaction}

\begin{figure*}[t]
\centering
\includegraphics[width = 180 mm]{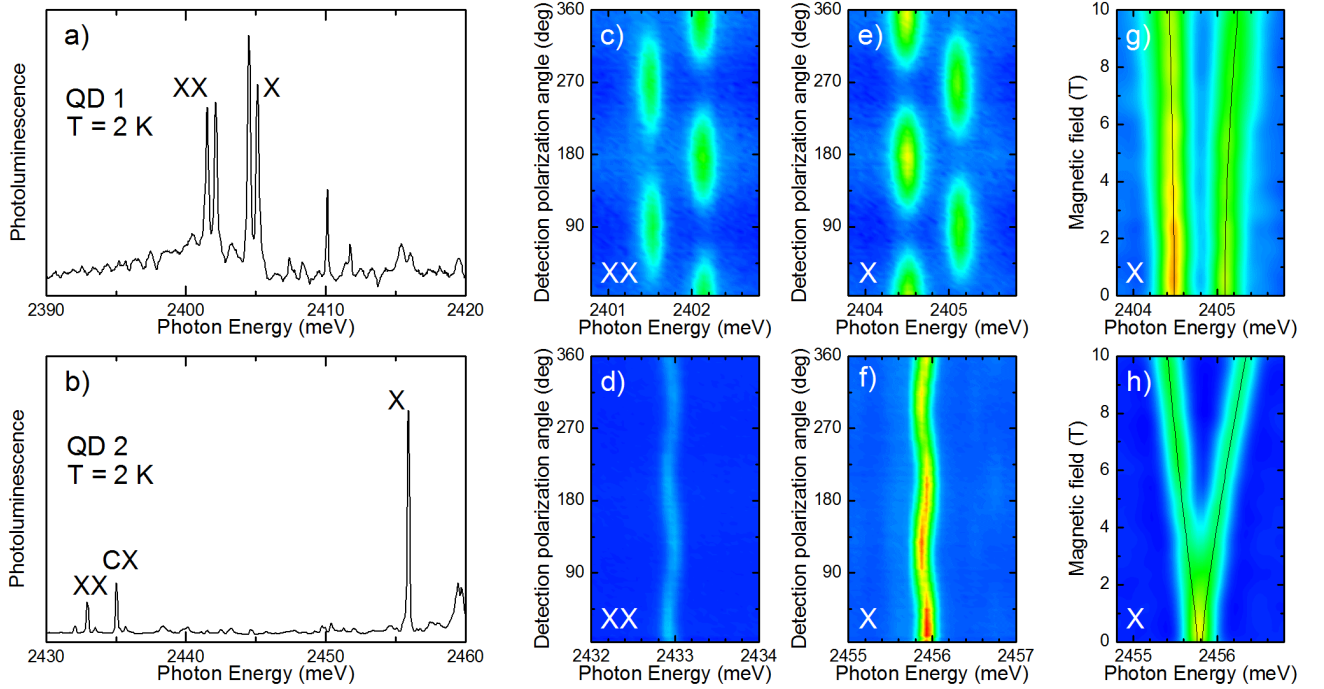}
\caption{Exemplary single Cd(Se,Te)/ZnSe QDs from the sample produced by the Mn-assisted growth: a) and b) polarization integrated emission spectra with indicated X, XX and charged exciton (CX) transitions, c) and d) emission spectra \emph{vs} detected linear polarization angle for the XX, e) and f) emission spectra \emph{vs} detected linear polarization angle for the X, g) and h) energy dependence of the X transition on magnetic field applied in Faraday configuration, for QD1 and QD2, respectively. Simulation (black lines) fitted to the experimental data (see text) yields g-factors: $g = 0.95$ and $g = 1.6$ for the QD1 and the QD2, respectively.} \label{fig:exampleQDs}
\end{figure*}

The Figure~\ref{fig:exampleQDs} shows example emission spectra of two individual QDs (labeled as QD1 and QD2), both coming from the sample produced by the Mn-assited growth. The emission lines are identified as the X and XX recombination.\cite{Kudelski:JofLumin2005} The energy difference $\deltaE$ amounts to 2.99~meV and 20.38~meV in the case of the QD1 and the QD2, respectively. As discussed in the previous Section, the $\deltaE$ acts as a measure of the \emph{e-h} wavefunction overlap. Thus, the QD1 and QD2 represent two extreme cases of a "small" and "large" \emph{e-h} overlap, respectively. Note that the QD1 exhibits much lower intensity as compared to the QD2, as expected for a QD with a decreased exciton oscillator strength resulting from a decreased \emph{e-h} overlap.\cite{Nirmal:PRL1995, Sujanah:OptCommun2015}

The spectra of the XX and X transitions from the QD1 and the QD2 plotted as a function of a detected linear polarization angle reveal two linearly polarized orthogonal components (see panels c-f) in the Fig.~\ref{fig:exampleQDs}). An exact antiphase of two components of the X doublet with respect to the XX doublet confirms that the transitions indeed originate from the same QD.\cite{Kudelski:JofLumin2005} The fine structure splitting $\delta_1$ for the X and XX from the QD1 amounts to 0.61~meV. In the case of the QD2, much smaller $\delta_1$ equal to 0.07~meV is found.

\begin{figure}[t]
\centering
\includegraphics[width= 87 mm]{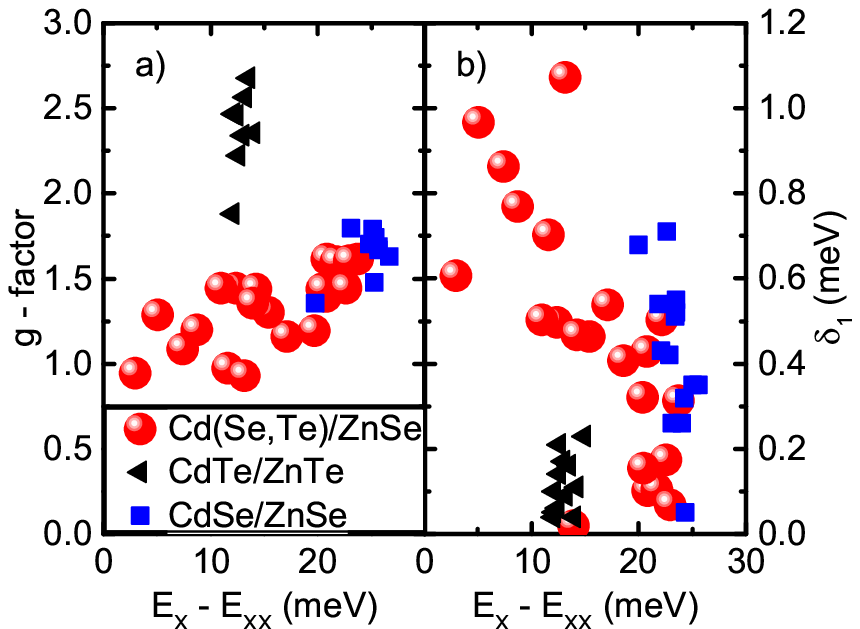}\\
\caption{Exciton parameters describing Cd(Se,Te)/ZnSe QDs produced by the Mn-assisted growth: a) Land{\'e} factor and b) fine-structure splitting $\delta_1$, plotted as a function of exciton - biexciton energy difference $\deltaE$. Respective values for CdSe/ZnSe and CdTe/ZnTe QDs are also shown for a reference (after Ref.~\onlinecite{Kobak:unpublished}).} \label{fig:stats}
\end{figure}

Figures.~\ref{fig:exampleQDs}g) and h) show evolution of the X transition in the magnetic field applied in the Faraday configuration. The black lines represent the simulation of a Zeeman splitting of the exciton ground state in an anisotropic QD following the equation $E_{\mathrm{X}}(B) = E_0 \pm \frac{1}{2}\sqrt{\delta_1^2 + \left(g \mu_B B\right)^2}$, where $E_0$ represent transition energy at $B=0$~T, $\delta_1$ is the X fine-structure splitting, while $g$ and $\mu_B$ represent parallel excitonic Land{\'e} factor and Bohr magneton, respectively. The g-factor determined from the fit equals to 1.61 for the QD2, which is quite typical value for CdSe/ZnSe QDs. In contrast, the $g = 0.95$ found for the QD1 is atypically small as for the QD based on any selenium or tellurium compound. The parameters describing the X emission determined in this Subsection for the QD1 and QD2 suggest that in the studied QDs the $\delta_1$ decreases, while the $g$ increases with the increasing $\deltaE$. The results obtained on a statistics of around twenty QDs are discussed in the next Subsection.

%#################################################################
\subsection{\label{sec:Discussion} Discussion}

The values of the excitonic Land{\'e} factor and the $\delta_1$ determined for a number of Cd(Se,Te)/ZnSe QDs are plotted against the $\deltaE$ in Fig.~\ref{fig:stats}a). The values for typical, binary type CdTe/ZnTe and CdSe/ZnSe QDs are also presented for a reference (after Ref.~\onlinecite{Kobak:unpublished}).

A striking property of the studied QDs is that the $\deltaE$ covers a very wide energy range from 3~meV to 25~meV (see Fig.~\ref{fig:tau} and Fig.~\ref{fig:stats}). In the case of CdSe or CdTe based QDs the $\deltaE$ exhibits typically a low scatter and remains in a relatively narrow range of 20-25~meV\cite{Kulakovskii:PRL1999, Bacher:PRL1999, Puls:PRB1999, Patton:PRB2003, Kobak:NatureCom2014, Piwowar:JofLumin2016, Smolenski:NatureCom:2016, Kobak:unpublished} or 12-15~meV\cite{Besombes:PRB2002, Kazimierczuk:PRB2011, Kobak:NatureCom2014, Sujanah:OptCommun2015, Kobak:unpublished}, respectively. Hence, the $\deltaE$ in our case spans from values typical for the CdSe/ZnSe QDs, through values typical for the CdTe/ZnTe QDs, down to much lower values, which were not observed before for the CdSe or the CdTe based QDs systems and which would not be expected for the Cd(Se,Te)/ZnSe QDs if they were homogeneous in a volume. The strongly reduced values of the $\deltaE$ mean that the \emph{e} and \emph{h} wavefunctions are separated in space much more than in pure CdSe or CdTe based QDs. It results most likely from a localization of the electrons in CdSe and the holes in CdTe rich regions (see Fig.~\ref{fig:bands}) present in the QD volume. We link the presence of such regions to the QD layer growth procedure (Sec.~\ref{sec:samples}), namely the formation of the QDs out of the deposited consecutively CdSe and CdTe layers. A possible strain would widen the CdTe bandgap and narrow the ZnSe bandgap, even further enhancing the carrier localization effects.\cite{Smith:NatureNano2008}

Moreover, the Fig.~\ref{fig:stats} reveals a clear increase of the excitonic Land{\'e} factor and a decrease of the $\delta_1$ with the increasing $\deltaE$. Such dependencies have not been reported for epitaxial II-VI QDs so far. As it is seen in Fig.~\ref{fig:stats}a), the g-factor attains the value as small as 0.95 for the QD characterized by the $\deltaE$ = 2.99~meV. At the first sight it is surprising, since typical g-factors of CdSe and CdTe QDs are both equal or larger than 1.6, what would result in the g-factor value exceeding 1.6 for a QD made of fully mixed (Cd,Se)Te. This is apparently not the case, which gives a hint that the decreased $\deltaE$ indeed reflects the increased \emph{e-h} separation resulting from a presence of CdTe and CdSe rich regions in the studied QDs. For the Cd(Se,Te)/ZnSe QDs with large values of the $\deltaE$, the excitonic g-factor is approximately the same as for the pure CdSe/ZnSe QDs (see Fig.~\ref{fig:stats}a)), that is around 1.6.\cite{Kulakovskii:PRL1999, Puls:PRB1999} The increase of the parallel Land{\'e} factor with the decreasing size of the confining potential in the growth direction, was previously observed in the case of QWs\cite{Hannak:SSC1995, Zhao:APL1996} and arsenide based QDs\cite{Kotlyar:PRB2001, Prado:PRB2004, Pryor:PRL2006, Kleemens:PRB2009}, as well as colloidal PbS nanocrystals.\cite{Turyanska:PRB2010} The effect was attributed to a quenching of the orbital momentum of a heavy hole when the hole wavefunction undergoes squeezing in the direction parallel to the magnetic field.\cite{Prado:PRB2004}

The $\delta_1$ varies with the increasing $\deltaE$ from around 1~meV, the value larger than for reference CdSe/ZnSe or CdTe/ZnTe QDs, down to the values below 0.06~meV (see Fig.~\ref{fig:stats}b)). It has been shown previously that a dominant contribution to the $\delta_1$ comes from the long-range component of the \emph{e-h} exchange interaction.\cite{Takagahara:PRB1993, Franceschetti:PRB1998, Takagahara:PRB2000, Bayer:PRB2002} The magnitude of the $\delta_1$, similarly as the magnitude of the short-range component, have been shown to increase with a decreasing QD size (thus with the increasing \emph{e-h} overlap).\cite{Takagahara:PRB1993, Franceschetti:PRB1998, Takagahara:PRB2000, Bayer:PRB2002} Thus, in the presently studied QDs, one should expect enhancement of the $\delta_1$ with the $\deltaE$. This is, however, not the case, what indicates that in the studied QDs, not just the overall \emph{e} and \emph{h} wavefunctions overlap, but specifically the anisotropy of the exciton wavefunction in the QD plane provides a main contribution to the $\delta_1$. A decrease of $\delta_1$ with a decreasing QD size was previously observed in the case of epitaxially grown InAs/GaAs QDs \cite{Seguin:PRL2005, Musial:SSC2012} and attributed to the dominating influence of piezoelectricity or heavy hole-light hole subbands mixing effects. In our case, the increase of the $\delta_1$ is correlated rather with the increasing Te content in the Cd(Se,Te)/ZnSe QDs, which translates into a larger respective spatial shift of the \emph{e} and \emph{h} in the QD plane.

We note that for the Cd(Se,Te)/ZnSe QDs with large values of the $\deltaE$ not only the g-factor, but also the fine structure splitting $\delta_1$ (0.1 - 0.3 meV) matches the one of the pure CdSe/ZnSe QDs\cite{Kulakovskii:PRL1999, Puls:PRB1999} (see Fig.~\ref{fig:stats}b)). This points toward binary type QD composition in the case of these QDs. This results most likely from an efficient substitution of Te atoms by Se atoms during the QD growth, due to a much higher sticking coefficient of the Se atoms as compared to the Te atoms.\cite{Yao:JCGrowth1978, Bonef:APL2015}

No systematic dependence of a diamagnetic coefficient $\gamma$ on the $\deltaE$ is found (not shown). The $\gamma$ is typically related to a span of the \emph{e} and \emph{h} wavefunctions in the QD plane. Thus, we state the \emph{e-h} wavefunctions overlap varies from dot to dot mainly in the sample growth direction.

%#################################################################
\section{Conclusions}

We have investigated systematically the impact of the electron-hole separation on optical properties of individual Cd(Se,Te)/ZnSe quantum dots. The reduced spectral density of QD emission lines obtained thanks to the Mn-assisted epitaxial growth has provided us an access to individual QDs emission. A statistics involving several individual QDs is collected. We have found that there exists a high degree of correlation between quantities being a direct measure of the \emph{e-h} wavefunctions overlap, such as the exciton radiative decay rate and the bright-dark exciton splitting $\delta_0$, with the exciton-biexciton energy difference $\deltaE$. The $\deltaE$ varies significantly from dot to dot, i.e., it changes from values typical for CdSe/ZnSe QDs and CdTe/ZnTe QDs, down to values as small as 3~meV, which were not reported previously for either of the two binary type systems. We attribute the wide distribution of the $\deltaE$ to a wide range of variation of the studied QDs composition and inhomogeneity affecting the \emph{e-h} wavefunctions overlap. In particular, the strongly reduced $\deltaE$ points toward a presence of the CdTe and CdSe rich regions in the studied QDs, which localize electron and hole, respectively, and account for an increased \emph{e-h} separation.

Moreover, in contrary to a simple expectation, the Cd(Se,Te)/ZnSe QDs g-factor is smaller than g-factors found previously in case of both, CdSe/ZnSe QDs and CdTe/ZnTe, binary type QD systems. Additionally, a distinct and so far not observed for II-VI QDs, dependencies of the g-factor and of the fine structure exchange splitting $\delta_1$ on the $\deltaE$ are found. The g-factor increases, while the $\delta_1$ decreases with the increasing $\deltaE$, that is with the increasing \emph{e-h} wavefunctions overlap.

Since an ability of selection of QDs with precisely defined emission properties is a prerequisite for QDs practical implementations, the demonstrated parametrization of the g-factor or the $\delta_1$ by the $\deltaE$ in the studied QDs provides a perspective for overcoming of the limits related to a scatter of the g-factor and the $\delta_1$ values observed typically in the QDs ensemble. At the same time, further efforts for obtaining QDs with a degree of separation of confined carriers controlled on the production stage are still desired.

\section*{Acknowledgments}

This work was supported by the Polish National Research Center projects DEC-2013/10/E/ST3/00215, DEC-2013/09/B/ST3/02603, DEC-2011/02/A/ST3/00131, DEC-2015/18/E/ST3/00559, and DEC-2011/02/A/ST3/00131, by the Polish Ministry of Science and Higher Education in years 2012-2016 as a research grant ``Diamentowy Grant'' and "Iuventus Plus" grant IP2014 034573. One of us (T.S.) was supported by the Foundation for Polish Science through the START programme. The work was carried out with the use of CePT, CeZaMat, and NLTK infrastructures financed by the European Union - the European Regional Development Fund within the Operational Programme "Innovative Economy". The use of the
FEI Tecnai Osiris TEM instrument located at the Facility for Electron Microscopy \& Sample Preparation of the University of Rzesz\'{o}w is acknowledged.

\section*{Appendix}

Determination of the average molar density $x$ of the Te atoms in the Cd(Se,Te) QDs bases on relative changes of a vertically integrated EDX signal. Formation of pure CdTe QDs would result in equal relative (with respect to the maximum determined for the barrier layer) decrease of Zn and Se signal in the QDs layer. As evidenced in Fig.~\ref{fig:HAADF}d), showing the vertically integrated EDX signal for of Zn, Se, Cd, Te, and Mn as a function of the distance from the QDs plane, this is, however, not the case. The different relative decrease of the Zn and Se signal in the QDs layer confirms the presence of the Se atoms in the QDs. Since a number of the Cd anions in the Cd(Se,Te) QDs is equal to the sum of the, substituting each other, Te and Se cations, the ratio of a relative decrease $\Delta I_{\mathrm{Se}}$ of the Se signal to a relative decrease $\Delta I_{\mathrm{Zn}}$ of the Zn signal (substituted by Cd) gives the ratio of the Te to Cd atom densities, thus the average molar density $x$ of the Te atoms in the Cd(Se,Te) QDs:
$$
x =  \frac{\Delta I_{\mathrm{Se}}}{\Delta I_{\mathrm{Zn}}}
$$

\begin{figure}[t]
\centering
\includegraphics[width = 87 mm]{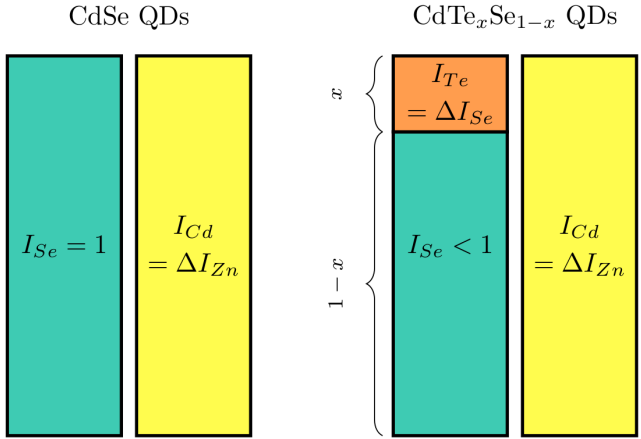}
\caption{Determination of average Te content in the Cd(Se,Te) QDs: the ratio of the Te to Cd atom densities, that is the average molar density $x$ of the Te in the Cd(Se,Te) QDs is set by the ratio of a relative decrease $\Delta I_{\mathrm{Se}}$ of the EDX signal of Se to a relative decrease $\Delta I_{\mathrm{Zn}}$ of the EDX signal of Zn (which is substituted by Cd).}
\end{figure}

We note that the EDX signal detection sensitivity depends in unknown way on a given element. As a result, the absolute values of EDX signal for each element are also not known. It is known, however, that the signals for Se and Zn in the ZnSe barrier region attain their maximum values. These values serve as a reference in determination of relative changes of the EDX signal in the QD layer for respective elements.

\bibliographystyle{apsrev_my}
%\bibliography{MS_CdSeTe_ZnSe_bibliography}

\end{document}